\documentclass[sigconf,natbib=true]{acmart}
\newcommand{\veca}{\textbf{a}}
\newcommand{\vecs}{\textbf{s}}
\newcommand{\vecg}{\textbf{g}}
\usepackage{array}
\usepackage{amsthm}
\usepackage{mathtools}
\usepackage{hyperref}
\usepackage{amsmath,amsfonts,bm}
\usepackage{subcaption,multirow,cleveref}
\crefname{algocf}{alg.}{algs.}
\Crefname{algocf}{Algorithm}{Algorithms}
\AtBeginDocument{%
  \providecommand\BibTeX{{%
    \normalfont B\kern-0.5em{\scshape i\kern-0.25em b}\kern-0.8em\TeX}}}

\setcopyright{none}
\copyrightyear{2018}
\acmYear{2018}
\acmDOI{XXXXXXX.XXXXXXX}

\acmConference[KDD '24]{ACM KDD 24}{August 25--29,
  2024}{Barcelona, Spain}
\acmPrice{15.00}
\acmISBN{978-1-4503-XXXX-X/18/06}

\usepackage[ruled,linesnumbered]{algorithm2e}
\newcommand{\nonl}{\renewcommand{\nl}{\let\nl\oldnl}}
\newtheorem{theorem}{Theorem}
\begin{document}


\title{Maximum-Entropy Regularized Decision Transformer with Reward Relabelling for Dynamic Recommendation}
\author{Xiaocong Chen}
\affiliation{%
  \institution{Data 61, CSIRO}
  \city{Eveleigh}
  \country{Australia}
}
\email{xiaocong.chen@data61.csiro.au}
\author{Siyu Wang}
\affiliation{%
  \institution{The University of New South Wales}
  \city{Sydney}
  \country{Australia}}
\email{siyu.wang5@unsw.edu.au}

\author{Lina Yao}
\affiliation{%
  \institution{Data 61, CSIRO}
  \city{Eveleigh}
  \country{Australia}}
\affiliation{
  \institution{The University of New South Wales}
  \city{Sydney}
  \country{Australia}}
\email{lina.yao@data61.csiro.au}

\renewcommand{\shortauthors}{Chen et al.}

\begin{abstract}
Reinforcement learning-based recommender systems have recently gained popularity. However, due to the typical limitations of simulation environments (e.g., data inefficiency), most of the work cannot be broadly applied in all domains. To counter these challenges, recent advancements have leveraged offline reinforcement learning methods, notable for their data-driven approach utilizing offline datasets. A prominent example of this is the Decision Transformer. Despite its popularity, the Decision Transformer approach has inherent drawbacks, particularly evident in recommendation methods based on it.  This paper identifies two key shortcomings in existing Decision Transformer-based methods: a lack of stitching capability and limited effectiveness in online adoption. In response, we introduce a novel methodology named Max-Entropy enhanced Decision Transformer with Reward Relabeling for Offline RLRS (EDT4Rec). Our approach begins with a max entropy perspective, leading to the development of a max-entropy enhanced exploration strategy. This strategy is designed to facilitate more effective exploration in online environments. Additionally, to augment the model’s capability to stitch sub-optimal trajectories, we incorporate a unique reward relabeling technique. To validate the effectiveness and superiority of EDT4Rec, we have conducted comprehensive experiments across six real-world offline datasets and in an online simulator.
\end{abstract}






\maketitle

\section{Introduction}

Reinforcement Learning (RL)-based Recommender Systems (RS) have emerged as powerful tools in a variety of domains, ranging from e-commerce~\cite{hu2018reinforcement,cai2018reinforcement} and advertising~\cite{cai2017real} to streaming services. Their effectiveness is particularly pronounced in dynamic environments, where users' interests are dynamic~\cite{chen2023deep}. In RLRS, an agent interacts with users by recommending items and receiving feedback. This feedback, typically in the form of user responses to recommendations, serves as rewards that inform the agent’s decisions. Utilizing this feedback loop, the agent continually refines its policy to better align with user preferences, ultimately aiming to maximize long-term rewards. This goal often translates into enhancing user click with the system. Despite their proven utility, RLRS frequently suffers data inefficiency, a challenge inherent to the interaction-centric nature of RL algorithms. This inefficiency arises as the systems must learn from limited user interactions, making the process of policy improvement slower and less robust. 

Recent studies have proposed using offline RL to address the challenges faced by RL-based Recommender Systems (RS)~\cite{Wang_2023,zhao2023user,chen2023opportunities}. Known as data-driven RL, offline RL~\cite{levine2020offline} leverages extensive offline datasets for the preliminary training of agents, allowing for the utilization of pre-existing data to train RLRS agents. A notable implementation is CDT4Rec~\cite{Wang_2023}, which integrates the Decision Transformer(DT)~\cite{chen2021decision} as its foundational structure, coupled with a causal mechanism for reward estimation. Similarly, DT4Rec, proposed by~\citet{zhao2023user}, utilizes DT to focus on capturing user retention, implementing an efficient reward prompting method for RLRS.

\begin{figure}
    \centering
    \includegraphics[width=\linewidth]{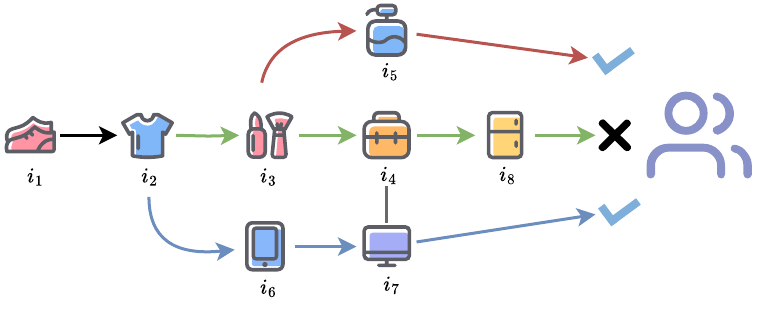}
    \caption{An example demonstrates that when DT is directly applied to RS will face the stitching problem (i.e., cannot learn from the sub-optimal trajectory). }
    \label{fig:stitching}
\end{figure}

However, directly applying the DT to RS still has several obstacles which require further investigation. The DT requires an offline dataset that contains sufficient \emph{expert} trajectories (i.e., those trajectories should be optimal and dense) that can cover almost all of the possibilities that the agent may face when interacting with the real environment. It will lead to two problems when applying the DT algorithm in RS. Firstly, in RS, the offline dataset is highly sparse~\cite{chen2022locality} which may not be able to generate enough expert trajectories. Under such a scenario, the DT needs to have the capability to learn from the sub-optimal trajectories. However, recent literature~\cite{li2023survey, zheng2022online,yamagata2023q,chen2023opportunities,nair2020awac} indicate that the vanilla DT lacks the capability of learning from sub-optimal trajectories (i.e., stitching). To better understand the stitching capability in DT, we have provided an illustration example under recommendation scenario in~\Cref{fig:stitching} to explain. Consider a user has three previous trajectories consisting of item click sequences: $(i_1,i_2,i_3,i_4,i_8)$, $(i_1,i_2,i_6,i_7)$, and $(i_1,i_2,i_3,i_5)$. In this illustration, each node represents an item. Notably, the trajectory $(i_1,i_2,i_3,i_4,i_8)$ receives a reward of 0 (i.e., the user does not click it), whereas a similar trajectory $(i_1,i_2,i_6,i_7)$ yields a positive reward. We categorize the 0 reward scenario as sub-optimal. Upon closer examination, we observe a linkage between items $i_4$ and $i_7$, suggesting the potential to create a more rewarding trajectory by combining elements from $(i_1,i_2,i_3,i_4,i_8)$ and $(i_1,i_2,i_6,i_7)$. For instance, a newly synthesized trajectory $(i_1,i_2,i_3,i_4,i_7)$ is predicted to achieve a positive reward. Our reward relabeling strategy addresses this by assigning weights to each node, ensuring the agent avoids less rewarding paths like reaching node $i_8$.

Moreover, since the DT believes that all of the possibilities are covered in the offline dataset, the exploration will be abandoned when fine-tuning the online environments~\cite{li2023survey}. It will be crucial when applying to the RS, since the users' interests are dynamic and rapidly changing, and the recorded offline dataset can not reflect users' intentions completely. The agent still needs to conduct certain steps of exploration to collect trajectories as suggested by recent works~\cite{chen2021exploration,chen2023deep}. These limitations hinder the agent's performance, particularly when working with sub-optimal offline datasets that do not fully encapsulate the range of possible user actions, a situation more akin to real-world scenarios.

To address those two challenges, in this study, we introduce a novel model, named Max-Entropy enhanced Decision Transformer with Reward Relabeling for Offline RLRS (EDT4Rec), to overcome the identified challenges in RL-based Recommender Systems. Drawing inspiration from the Soft Actor-Critic (SAC) approach~\cite{haarnoja2018soft}, EDT4Rec incorporates the concept of max-entropy exploration. It also introduces an innovative reward relabeling strategy, whereby each node in a trajectory is assigned a reward to foster the generation of more optimal trajectories and facilitate the learning of a more effective recommendation policy\footnote{For clarity, we will use "policy" as an abbreviation in subsequent sections.}.
Our contributions in this work are threefold:
\begin{itemize}
\item We propose EDT4Rec, a novel model that integrates max-entropy exploration and a unique reward relabeling strategy, enhancing the offline RLRS framework.
\item To address the stitching problem in DT, we design a novel reward relabeling strategy. Moreover, we design a new max-entropy exploration mechanism to enable the agent can conduct the exploration when fine-tuning on the online environments.
\item We validate the efficacy of EDT4Rec through extensive experiments on six public datasets and in an online simulation environment, demonstrating its superiority over existing methods.
\end{itemize}

\section{Problem Formulation}
The recommendation problem can be conceptualized as an agent striving to achieve a specific goal through learning from user interactions, such as item recommendations and subsequent feedback. This scenario is aptly formulated as a RL problem, where the agent is trained to interact within an environment, typically described as a Markov Decision Process (MDP)~\cite{chen2023deep}. 
The components of an MDP are represented as a tuple $(\mathcal{S},\mathcal{A}, \mathcal{P}, \mathcal{R}, \gamma)$ where,
\begin{itemize}
    \item State $\mathcal{S}$: The state space, with $s_t \in \mathcal{S}$ representing the state at timestep $t$, which normally contain users' previous interest, users' demographic information etc.
    \item Action $\mathcal{A}$: The action space, where $a_t \in \mathcal{A}(s_t)$. $a_t$ is the action taken when given a state $s_t$. It normally refers to recommended items.
    \item Transition Probability $\mathcal{P}$: denoted as $p(s_{t+1}|s_t, a_t) \in \mathcal{P}$, is the probability of transitioning from $s_t$ to $s_{t+1}$ when action $a_t$ is taken.
    \item Reward $\mathcal{R}$: $\mathcal{S} \times \mathcal{A} \to \mathbb{R}$ is the reward distribution, where $r(s, a)$ indicates the reward received for taking action $a$ when observing the state $s$.
    \item Discount-rate Parameter $\mathcal{\gamma}$: $\mathcal{\gamma} \in [0, 1]$ is the discount factor which use to balance previous reward and immediate reward.
\end{itemize}

In this context, an RL agent follows its policy $\pi$, a mapping from states to actions to make recommendations. The objective of RL is expressed as:
\begin{align*}
    \mathbb{E}_{\tau} \bigg[\sum_{t=0}^{\infty} \gamma^k r(s_t, a_t)\bigg]
\end{align*}

\emph{Offline} reinforcement learning diverges from traditional RL by exclusively utilizing pre-collected data for training, without the necessity for further online interaction~\cite{levine2020offline}. In offline RL, the agent is trained to maximize total rewards relies on a static dataset of transitions $\mathcal{D}$ for learning. This dataset could comprise previously collected data or human demonstrations. Consequently, the agent in offline RL lacks the capability to explore and interact with the environment for additional data collection. The dataset $\mathcal{D}$ in an offline RL-based RS can be formally described as $\{(s_t^u, a_t^u, s_{t+1}^u, r_t^u)\}$, adhering to the MDP framework $(\mathcal{S},\mathcal{A}, \mathcal{P}, \mathcal{R}, \gamma)$. For each user $u$ at timestep $t$, the dataset includes the current state $s_t^u\in\mathcal{S}$, the items recommended by the agent via action $a_t^u$, and the user feedback $r_t^u$.

\section{Related Work}
\noindent\textbf{RL-based Recommender Systems.}
Reinforcement learning-based recommendation methods approach the interactive process of making recommendations as a MDP~\cite{chen2023deep}. This approach can be categorized into two primary branches: model-based and model-free methods. In the realm of model-based techniques,~\citet{bai2019model} introduced a method employing generative adversarial training to simultaneously learn the user behavior model and update the recommendation policy. Recently, there has been a noticeable shift in the literature towards embracing model-free techniques for reinforcement learning-based recommendations. Notably,~\citet{zheng2018drn} pioneered the introduction of RL to enhance news recommender systems, and~\citet{zhao2018deep} further extended its application to the page-wise recommendation scenario.
Both of these approaches employ Deep Q-Networks (DQN)~\cite{mnih2013playing} to encode user and item information, effectively improving the quality of news recommendations. Moreover,~\citet{chen2020knowledge} incorporated knowledge graphs into the reinforcement learning framework, resulting in enhanced decision-making efficiency. Additionally,~\citet{chen2021generative} introduced a novel generative inverse reinforcement learning approach for online recommendations. This method autonomously extracts a reward function from user behavior, further advancing state-of-the-art in online recommendation systems.

\vspace{1mm}\noindent\textbf{Offline RL in RS.}
Recent studies have started exploring the prospect of integrating offline RLRS~\cite{chen2023opportunities}. Notably,~\citet{Wang_2023} introduced a novel model called the Causal Decision Transformer for RS (CDT4Rec). This model incorporates a causal mechanism designed to estimate the reward function, offering promising insights into the offline RL-RS synergy. Similarly,~\citet{zhao2023user} introduced the Decision Transformer for RS (DT4Rec), utilizing the vanilla Decision Transformer as its core architecture to provide recommendations, demonstrating its potential in the field.
Furthermore,~\citet{gao2023alleviating} delved into the examination of the Matthew effect within offline RL in Recommender Systems. This work contributes to our understanding of how offline RLRS can mitigate or address this phenomenon.

\section{Methodology}
\subsection{Decision Transformer and Overall Model Architecture}
In this work, we proposed a max-Entropy enhanced Decision Transformer with reward relabeling for offline RLRS(EDT4Rec).
Our proposed EDT4Rec leverages the DT framework~\cite{chen2021decision}, a pivotal element in recent advancements in offline RLRS, as seen in works like~\cite{Wang_2023,chen2023opportunities,zhao2023user}
The Decision Transformer processes a trajectory, denoted as $\tau$, by treating it as a sequence comprising three distinct types of input: return-to-go (RTG), states, and actions, represented as  $(g_1,s_1,a_1,\cdots, g_{|\tau|}, s_{|\tau|}, a_{|\tau|})$. Specifically, the initial RTG $R_1$, corresponds to the trajectory's total return, calculated as $\sum_{i=0}^{|\tau|} r_i$. At each timestep $t$, DT utilizes tokens from the most recent $K$ timesteps to generate an action $a_t$. This $K$ is a hyperparameter, defines the context length the transformer uses for its computations. The deterministic policy learned by DT, denoted as $\pi_{DT}(a_t|\vecs_{K,t},\vecg_{K,t})$, where $\vecs_{K,t}$ is shorthand for the sequence of $K$ past states $\vecs_{\max(1,t-K+1):t}$ and similarly for $\vecg_{K,t}$. A crucial aspect of DT in this context is the application of a causal mask to predict the sequence of actions. 

To simplify the following analysis, we postulate that the data distribution (i.e., for the dataset), denoted as $\mathcal{T}$,   produces subsequences of actions, states, and RTG values, each of length $K$, originating from the same trajectory. 
To facilitate our explanation, we slightly diverge from standard notation and represent a  sample from $\mathcal{T}$ as $(\veca,\vecs,\vecg)$ where all the vectors contain subelements. This notation choice simplifies the presentation of our training objective.
The core of DT training strategy involves instructing the policy to accurately predict action tokens. This prediction adheres to the standard $l_2$ loss, which is mathematically represented as:
\begin{align}
    \mathbb{E}_{(\veca,\vecs,\vecg)\sim\mathcal{T}}\bigg[\frac{1}{K}\sum_{k=1}^K\big(a_k-\pi_{DT}(\vecs_{K,k},\vecg_{K,k})\big)^2\bigg]. \label{eq:dt}
\end{align}
This equation encapsulates the expectation of the mean squared difference between the actual action tokens and those predicted by the DT policy, averaged over all $K$ elements in the subsequence.
In practice, to implement this approach within the framework of an offline dataset $\mathcal{D}$, we uniformly sample these length-$K$ subsequences.
\begin{figure*}[h]
  \centering
  \includegraphics[width=\linewidth]{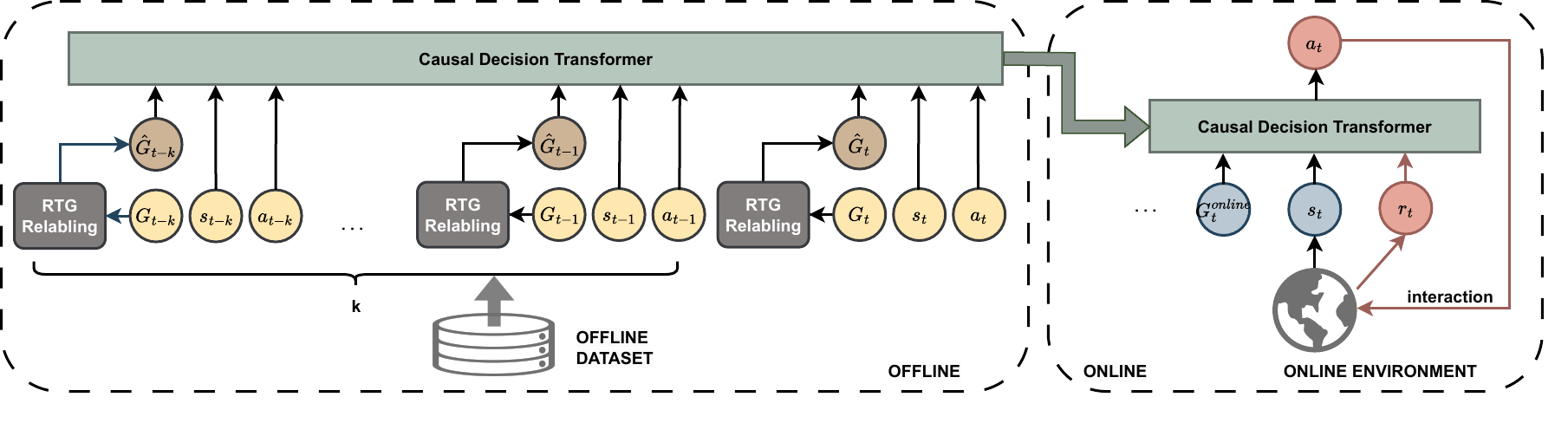}
  \caption{The overall structure of the proposed EDT4Rec. The backbone is the causal decision transformer.}
  \label{fig:overrall}
\end{figure*}

In EDT4Rec, we have introduced the following two key modifications: max-entropy enhanced Exploration and RTG relabeling.
The existing approach of training policies in decision-transformer-based offline RLRS~\cite{Wang_2023,chen2023opportunities} on purely offline datasets often encounters a critical limitation: the trajectories in these datasets usually do not yield high returns and cover only a restricted portion of the state space. This limitation stems from two parts: i). the potential distribution difference between the offline dataset and the online environment and, ii). the sub-optimality of the training data available offline. A straightforward strategy to mitigate the distribution difference is the fine-tuning of pre-trained RL agents through online interactions. However, the learning formulation of a standard decision transformer does not naturally lend itself to effective online learning. Moreover, in scenarios with sub-optimal trajectories, the lack of informative rewards (often zero) hinders effective policy updates.

\subsection{Max-Entropy Enhanced Exploration}
In the initial phase of our methodology, we propose a probabilistic learning objective, intended to augment exploration within the standard DT framework. This setup is geared towards developing a stochastic policy, with the primary goal of maximizing the likelihood of the observed dataset. For contexts involving continuous action spaces, we opt for the conventional approach of using a multivariate Gaussian distribution with a diagonal covariance matrix. This distribution models the action probabilities, conditional upon states and RTG values. The policy parameters are represented by $\theta$,
\begin{align}
    \pi_\theta(a_t|\vecs_{K,t},\vecg_{K,t}) = \mathcal{N}\Big(\mu_\theta(\vecs_{K,t},\vecg_{K,t}),\sum_\theta (\vecs_{K,t},\vecg_{K,t})\Big), \forall t\in T,
\end{align}
where $\sum_\theta$ denotes the diagonal covariance matrix.

Consider a stochastic policy, we aim to maximize the log-likelihood of the trajectories present in our training dataset. This objective is equivalently achieved by minimizing the negative log-likelihood (NLL) loss. The loss function, $J(\theta)$, is defined as follows: 
\begin{align}
    J(\theta) & = \frac{1}{K}\mathbb{E}_{(\veca,\vecs,\vecg)\sim\mathcal{T}}[-\log\pi_\theta(\veca|\vecs,\vecg)] \notag \\
    & =\frac{1}{K}\mathbb{E}_{(\veca,\vecs,\vecg)\sim\mathcal{T}}\bigg[\sum_{k=1}^K-\log\pi_\theta(a_k|\vecs_{K,k},\vecg_{K,k})\bigg]. \label{eq:newdt}
\end{align}
It's important to note that the policies considered here encompass the deterministic policies utilized by the standard DT. Optimizing~\Cref{eq:dt} is effectively equivalent to optimizing~\Cref{eq:newdt}, under the assumption that the covariance matrix $\sum_\theta$ is diagonal and the variances remain constant across all dimensions. This scenario represents a special case encompassed by our assumption.

A crucial aspect of an online RL algorithm is its capacity to balance exploration and exploitation. However, the traditional formulation of a DT, as seen in~\Cref{eq:newdt}, does not inherently facilitate exploration. To remedy this, we can use the policy entropy to encourage exploration, defined as:
\begin{align}
    H_\theta^\mathcal{T}[\veca|\vecs,\vecg] = & \frac{1}{K}\mathbb{E}_{(\vecs,\vecg)\sim\mathcal{T}}\big[H(\pi_\theta(\veca|\vecs,\vecg))\big] \notag \\
    =& \frac{1}{K}\mathbb{E}_{(\vecs,\vecg)\sim\mathcal{T}}\big[\sum_{k=1}^K H[(\pi_\theta(a_k|\vecs_{K,k},\vecg_{K,k})]\big], \label{eq:policyentropy}
\end{align}
where $H[\pi_\theta(a_k)]$ denotes the entropy of the distribution $\pi_\theta(a_k)$. This policy entropy is contingent on the data distribution $\mathcal{T}$, which remains static during the offline pretraining phase but becomes dynamic in the finetuning phase as it incorporates online data acquired through exploration. In line with max-entropy RL algorithms like SAC~\cite{haarnoja2018soft}, we explicitly impose a lower bound on the policy entropy to encourage exploration. We aim to solve the following constrained problem:
\begin{align}
    \min_\theta J(\theta) \quad \text{subject to } H_\theta^\mathcal{T}[\veca|\vecs,\vecg] \geq \beta, \label{eq:minobject}
\end{align}
where $\beta$ is a predetermined hyperparameter.  Practically, we address the dual problem of~\Cref{eq:minobject} to avoid the direct handling of the inequality constraint. Thus, we consider the Lagrangian,
\begin{align}
    \max_{\lambda\geq 0}\min_\theta  L(\theta,\lambda) \text{ where } L(\theta,\lambda) = J(\theta)+\lambda(\beta-H_\theta^\mathcal{T}[\veca|\vecs,\vecg]). \label{eq:primarygoal}
\end{align}
The solution involves alternating optimization of $\theta$ and $\lambda$. Specifically, optimizing $\theta$ with fixed $\lambda$ is  is achieved by,
\begin{align}
    \min_\theta J(\theta) - \lambda H_\theta^\mathcal{T}[\veca|\vecs,\vecg], \label{eq:optimizetheta}
\end{align}
and optimizing  $\lambda$ with a fixed $\theta$,
\begin{align}
    \min_{\lambda\geq 0} \lambda (H_\theta^\mathcal{T}[\veca|\vecs,\vecg]-\beta).\label{eq:optimizelambda}
\end{align}
Next, we will conduct a theoretical analysis to delineate how our EDT4Rec differs from SAC.

\Cref{eq:optimizetheta} represents the regularized form of our primal problem as outlined in~\Cref{eq:primarygoal}, where the dual variable $\lambda$ assumes the role of a temperature variable, a common element in many regularized RL formulations. A fundamental distinction of our approach from SAC and other traditional RL methods lies in our loss function $J(\theta)$, which is defined as the negative log-likelihood (NLL) rather than the discounted return. This shift signifies a focus on supervised learning of action sequences rather than explicitly aiming to maximize returns. Therefore, the objective in~\Cref{eq:optimizetheta} can be interpreted as minimizing the expected discrepancy between the log-probability of observed actions and the $\lambda$-scaled log-probability of actions sampled from $\pi_\theta(\cdot|\vecs,\vecg)$. In essence, we aim to train $\pi_\theta$ to approximate the observed action distribution, allowing for a controlled degree of deviation, with $\lambda$ explicitly regulating this mismatch.

It's important to note that $H_\theta^\mathcal{T}[\veca|\vecs,\vecg]$ is technically a cross conditional entropy. This arises because the training data distribution $\mathcal{T}$ is not identical to the action-state-RTG marginals induced by the current policy $\pi_\theta$ and the transition probability $\mathcal{P}$. In the offline training phase, $\mathcal{T}$ is a fixed offline data distribution, $\mathcal{T}_{offline}$, while during the online phase, $\mathcal{T}$ is represented through a replay buffer that stores online data and is dependent on the current policy $\pi_theta$ and previously gathered data. 

It is worth mentioning that our policy entropy, as defined in~\Cref{eq:policyentropy}, operates at the sequence level rather than at the transition level. This distinction leads to a significant difference in the constraints applied in our primal problem (\Cref{eq:primarygoal}) compared to those used in the SAC framework. For the sake of simplicity, if we momentarily set aside the RTG variable $\vecg$ in our policy entropy calculation, we can see a clear contrast. While SAC imposes a lower bound $\beta$ on the policy entropy at each timestep, our entropy $H_\theta^\mathcal{T}[\veca|\vecs,\vecg]$ is computed as an average over $K$ consecutive timesteps. 

\subsection{RTG Relabeling}
The reward conditioning approach prevalent in prior works typically involves conditioning on an entire trajectory sequence using the sum of rewards for that sequence (i.e.,$\sum_{i=0}^{|\tau|}r_i$). However, this method encounters difficulties in tasks that require stitching – the process of learning an optimal policy from sub-optimal trajectories by effectively combining them. By contrast, the Q-learning approach addresses this by independently propagating the value function backwards at each timestep using the Bellman backup. It integrates information for each state across different trajectories, thereby circumventing the stitching issue inherent in the reward conditioning approach. Our proposed solution to this challenge involves a novel approach: relabeling the RTG values using learned Q-functions. By applying this relabeling to the dataset, the reward conditioning approach can then effectively utilize optimal segments from within sub-optimal trajectories. This enhancement allows for the extraction and integration of valuable sub-trajectory segments, thus addressing the fundamental limitation of the original reward conditioning method in tasks requiring effective stitching of policy learning.

In addressing the relabeling of RTG values with learned Q-functions, it's important to recognize that simply substituting all RTG values with Q-function estimates is not always effective. This is particularly true in scenarios characterized by long time horizons and sparse rewards, where the accuracy of learned Q-functions can vary significantly. Our goal is to selectively replace RTG values in instances where the learned Q-functions provide reliable estimations. To achieve this, we incorporate the CQL framework~\cite{kumar2020conservative}, renowned for learning the lower bounds of value functions in offline Q-learning algorithms. The relabeling of RTG values is conducted selectively: it occurs when the RTG within a trajectory is lower than the lower bound estimated by CQL. This approach ensures that RTG values are replaced only when the learned value function is reasonably accurate or closer to the true value. Furthermore, our method extends the impact of this relabeling process through the trajectory. We apply reward recursion, defined as $R_{t-1} = r_{t-1}+R_t$, to propagate the revised RTG values to all preceding timesteps in the trajectory. This ensures that the influence of the adjusted RTG is not isolated but is instead integrated throughout the trajectory, enhancing the overall consistency and accuracy of our approach.

To implement our approach, we begin by initializing the RTG of the final state in a trajectory to zero, denoted as $R_T = 0$. Then, we proceed in reverse chronological order, starting from the end of the trajectory and moving backwards towards the initial state. The process involves a series of steps at each state in the trajectory. Firstly, for the current state $s_t$, we compute its state value using the learned value function, expressed as $\hat{V}(s_t)=\mathbb{E}_{a\sim\pi(a|s_t)}[\hat{Q}(s_t,a)]$, where $\pi$ represents the learned policy. This value function estimation reflects the expected value of actions taken in the current state according to the policy. Next, we compare this computed value function, $\hat{V}(s_t)$, with the existing RTG value at the same state (i.e., $R(t)$). If $\hat{V}(s_t)$ is greater than $R_t$, we update the RTG for the previous timestep $R_{t-1}$ to the sum of the reward at timestep $t-1$, $r_{t-1}$ and the estimated state value $\hat{V}(s_t)$. If not, $R_{t-1}$ is set to $r_{t-1}+R_t$. This relabeling process is repeated for each state in the trajectory until we reach the initial state. Through this backward iteration, we effectively update the RTG values based on the learned value function, ensuring that the RTG reflects a more accurate estimate of the expected returns from each state.

The relabeling process described earlier, while effective, can potentially disrupt the inherent consistency between rewards and RTG values in the input sequence of a Decision Transformer (DT). Ideally, the RTG value should always align with the sum of future rewards, adhering to the relationship $R_t = r_t+R_{t+1}$. However, the relabeling process may inadvertently violate this rule. To preserve this critical consistency within the DT input sequence, we adopt a method of regenerating the RTG values. For an input sequence $\{\hat{R}_{t-K+1},\cdots,\hat{R}_{t-1},\hat{R}_t\}$, we start by setting the last RTG $\hat{R}_t = R_t$. Then, we apply the formula $\hat{R}_{t'} = r'_t + \hat{R}_{t'+1}$ in a backward sequence until we reach $t'=t-K+1$. By doing so, we maintain the consistency between rewards and RTGs throughout the sequences. The comprehensive algorithm of RTG relabeling process can be found in~\Cref{alg:rtg_relabel}.

\begin{algorithm}[!ht]
\SetAlgoLined
 \SetKwInOut{Input}{input}
 \SetKwInOut{Output}{output}
 \Input{reward $r_{1:T}$, learned value function $\hat{V}(s)$, Trajectory length $T$, context length $K$}
 \Output{RTG $\hat{R}_{1:T}$}
    \tcp{Step 1. relabel RTG $R_T$}
    $R_T \leftarrow 0$\;
    $i \leftarrow T$\;
    \While{i > 0}{
        $R_{t-1} \leftarrow r_i + \max(R_t, \hat{V}(s_i)$\;
        $i\leftarrow i - 1$\;
    }
    \tcp{Step 2. relabel RTG $\hat{R}_T$ by considering the inconsistency}
    $\hat{R}_T \leftarrow R_t$\;
    $i \leftarrow T$\;
    \While{i > 0}{
        $\hat{R}_t \leftarrow r_i + \hat{R}_{i+1}$\;
        $i\leftarrow i - 1$\;
    }
     \caption{RTG Relabeling without inconsistency}
     \label{alg:rtg_relabel}
\end{algorithm}

A potential concern with our relabeling strategy is validating that the learned value function indeed represents the lower bound of the true value function, meaning it should be as close as possible to the optimal value function. To address this, we utilize the CQL~\cite{kumar2020conservative} to learn the Q-value which is defined as,
\begin{align}
    \hat{Q} \leftarrow & \arg\min_Q \alpha (\mathbb{E}_{s\sim\mathcal{D}, a\sim\mu(a|s)}[Q(s,a)]-\mathbb{E}_{s,a\sim\mathcal{D}}[Q(s,a)])\notag \\
    & + \frac{1}{2}\mathbb{E}_{s,a,s'\sim\mathcal{D}, a'\sim\mu(a'|s')}[(r(s,a)+\gamma\hat{Q}(s',a')-Q(s,a))^2]. \label{eq:cql}
\end{align}
Supporting this approach, we reference a theorem from the CQL framework (the proof of which can be found in the original paper):
\begin{theorem}[Lower Bound of CQL~\cite{kumar2020conservative}]
    The value of the policy under the Q-function from~\Cref{eq:cql}, $\hat{V}_\pi(\vecs) = \mathbb{E}_{\pi(\veca|\vecs)}[\hat{Q}_\pi(\vecs,\veca)]$, lower-bounds the true value of the policy obtained via exact policy evaluation.
\end{theorem}

In practice,  the optimal value function $Q^*(s,a)$  is unknown, compelling DT to rely on RTG as an alternative. RTG values, typically gathered through the behavior policy or policies, are often not optimal and tend to be significantly lower than their optimal value function counterparts, $Q^*(s,a)$ This discrepancy presents a challenge in accurately modeling the optimal policy. However, by employing CQL for learning the Q-function, we can effectively address this challenge. The aforementioned theorem from the CQL framework assures that the value function estimated under CQL acts as a reliable lower bound of the true policy value. Consequently, this theorem supports the premise that our relabeling process, guided by the learned Q-function, shifts the RTG values in the training dataset closer to the optimal value function.To further substantiate this claim, we also provide a straightforward proof. This proof demonstrates how the application of CQL in learning the Q-function ensures that the relabeled RTG values in our training dataset are more aligned with what would be expected from the optimal value function, thus enhancing the overall accuracy and effectiveness of the DT in policy modeling.

\subsection{Training Procedure}

We implement the aforementioned formulation through a transformer architecture, adapting and extending the existing DT framework. In EDT4RRec, we introduce modifications that align with the changes we've previously outlined. To predict the policy's mean and log variance, we utilize two separate fully connected layers at the output stage of the model. This architectural choice allows for a more nuanced and effective representation of the policy's characteristics.~\Cref{{alg:overall_training}} summarizes the overall training pipeline in the proposed method.

\begin{algorithm}[!ht]
\SetAlgoLined
 \SetKwInOut{Input}{input}
 \SetKwInOut{Output}{output}
 \Input{offline dataset $\mathcal{D}$, exploration RTG $\vecg_{online}$, reply buffer $\mathcal{D}_{re}$, number of round $i$, number of iteration $I$, batch size $B$, target policy parameter $\theta$}
    $\mathcal{D}_{re} \leftarrow $ top $N$ trajectories from $\mathcal{D}$ \;
    \For{round $1\cdots i$}{
        Trajectory $\tau \leftarrow \pi_\theta(\cdot|\vecs,\vecg_{online})$\;
        \tcp{if the $\mathcal{D}_{re}$ is full, the oldest trajectories will be removed. }
        $\mathcal{D}_{re}\leftarrow \mathcal{D}_{re} \cup \tau$ \; 
        Compute the trajectory sampling probability $p(\tau)=|\tau|/\sum_{\tau \in \mathcal{D}}|\tau|$\;
        \For{$t=1\cdots I$}{
            Sample $B$ trajectories from $\mathcal{D}_{re}$\;
            \For{each trajectory $\tau$}{
                $\vecg \leftarrow$ relabeling using~\Cref{alg:rtg_relabel}\;
                $(\veca,\vecs,\vecg)\leftarrow$ a length of $K$ sub-trajectory uniformly sampled from $\tau$\;
            }
            $\theta\leftarrow$ one gradient update using the sampled $(\veca,\vecs,\vecg)$\;
        }
    }
     \caption{Overall training algorithm}
     \label{alg:overall_training}
\end{algorithm}

In our implementation, both the online goal $\vecg_{online}$ and the context length $K$ are set to 2, a decision whose rationale and implications are further explored in our hyperparameter study.

\section{Experiments}
In this section, we report the outcomes of experiments that focus on
the following three main research questions:
\begin{itemize}
    \item \textbf{RQ1}: How does EDT4Rec compare with other DT-based methods and traditional deep RL algorithms in online recommendation environments and offline dataset environments?
    \item \textbf{RQ2}: How do the hyper-parameters affect the performance in the \emph{online} simulation environment? 
    \item \textbf{RQ3}: How does each component in EDT4Rec contribute to the final performance in the \emph{online} simulation environment?
    
\end{itemize}
We concentrate our hyper-parameters and ablation study (i.e., RQ2 and RQ3) on online simulation settings since they are more closely suited to real-world environments, whereas offline datasets are fixed and do not reflect users' dynamic interests.
\subsection{Datasets and Environments}
In this section, we evaluate the performance of our proposed algorithm, EDT4Rec, against other state-of-the-art algorithms, employing both real-world datasets and an online simulation environment. We first introduce six diverse, public real-world datasets from various recommendation domains for our offline experiments:
\begin{itemize}
\item \textbf{Kuairand-1k-15policies}~\cite{gao2022kuairand}: An unbiased sequential recommendation dataset sourced from recommendation logs of a video-sharing mobile app.
\item \textbf{LibraryThing}: This online service facilitates book cataloging for users~\footnote{https://www.librarything.com/}. It is notable for its inclusion of social relationships between users, making it ideal for studying social recommendation algorithms.
\item \textbf{MovieLens}: We utilize the MovieLens-20M dataset\footnote{https://grouplens.org/datasets/movielens/20m/}, a widely recognized benchmark in recommender system research.
\item \textbf{GoodReads}: Sourced from the book review website \textit{GoodReads}, this dataset, compiled by~\cite{wan2018item}, includes various user interactions like ratings and reviews.
\item \textbf{Netflix}: Originating from the Netflix Prize Challenge\footnote{https://www.kaggle.com/datasets/netflix-inc/netflix-prize-data}, this well-known dataset primarily consists of user ratings.
\item \textbf{Book-Crossing}: Similar to MovieLens, this book-related dataset by~\cite{ziegler2005improving} focuses on rating information.
\end{itemize}

To  evaluate EDT4Rec's performance, we transformed the offline datasets into simulation environments that allow for interactive reinforcement learning experiences. This conversion aligns with methodologies outlined in existing research~\cite{chen2020knowledge,Wang_2023,zhao2023user}. Specifically, user interactions are converted into binary click behaviors to construct trajectories. For datasets containing ratings, any rating exceeding 75\% of the maximum rating is considered positive feedback, while the rest are treated as negative. The evaluation process adheres to the standards established in previous works~\cite{Wang_2023}.

In addition, we also experiment on a real online simulation platform to validate the proposed method. We use VirtualTB~\cite{shi2019virtual} as the major online platform in this work. In terms of evaluation metrics, we use the click-through rate (CTR) for the online simulation platform as the CTR is one of the built-in evaluation metrics of the VirtualTB simulation environment. 
For offline dataset evaluation, we employ a variety of evaluation metrics, including recall, precision, and normalized discounted cumulative gain (nDCG). 

\subsection{Baselines}
\begin{table*}[ht]
\caption{The overall results of our model comparison with several state-of-the-art models on different datasets. The highest results are in bold and significant under the 95\% confidence interval}\smallskip
\begin{minipage}[ht]{0.95\linewidth}
\resizebox{\columnwidth}{!}{%
\begin{tabular}{ccccccc}
\hline
\multicolumn{1}{c|}{Dataset} & \multicolumn{3}{c|}{Kuairand-1k} & \multicolumn{3}{c}{Librarything} \\ \hline
\multicolumn{1}{c|}{Measure (\%)} & \multicolumn{1}{c|}{Recall} & \multicolumn{1}{c|}{Precision} & \multicolumn{1}{c|}{nDCG} & \multicolumn{1}{c|}{Recall} & \multicolumn{1}{c|}{Precision} & nDCG \\ \hline
\multicolumn{1}{c|}{DDPG} & 29.441 $\pm$ 0.321 & 28.774 $\pm$ 0.233 & \multicolumn{1}{c|}{29.773 $\pm$ 0.298} &  12.128 $\pm$ 0.241 & 12.451 $\pm$ 0.242 & 13.925 $\pm$ 0.252\\ 
\multicolumn{1}{c|}{SAC} & 29.204 $\pm$ 0.292 & 28.162 $\pm$ 0.205 & \multicolumn{1}{c|}{29.455 $\pm$ 0.201} & 12.028 $\pm$ 0.237 & 12.311 $\pm$ 0.255 & 13.125 $\pm$ 0.204\\ 
\multicolumn{1}{c|}{TD3} & 29.115 $\pm$ 0.281 & 27.989 $\pm$ 0.247 & \multicolumn{1}{c|}{29.107 $\pm$ 0.188} & 11.943 $\pm$ 0.205 & 12.018 $\pm$ 0.278 & 12.878 $\pm$ 0.218\\  
\multicolumn{1}{c|}{DT} &  30.082 $\pm$ 0.212 & 29.951 $\pm$ 0.215 & \multicolumn{1}{c|}{30.134 $\pm$ 0.160} & 14.225$\pm$ 0.147 & 13.967 $\pm$ 0.207 & 13.987 $\pm$ 0.102\\    
\multicolumn{1}{c|}{DT4Rec} & 30.104 $\pm$ 0.233 & 29.989 $\pm$ 0.201 & \multicolumn{1}{c|}{30.156 $\pm$ 0.145} & 14.333$\pm$ 0.156 & 13.984 $\pm$ 0.224 & 14.024 $\pm$ 0.146\\  
\multicolumn{1}{c|}{CDT4Rec} & 30.322 $\pm$ 0.208 & 30.014 $\pm$ 0.178  & \multicolumn{1}{c|}{30.525 $\pm$ 0.168} & 15.229 $\pm$ 0.128 & 14.020 $\pm$ 0.201 & 14.768 $\pm$ 0.176 \\ 
\hline
\multicolumn{1}{c|}{EDT4Rec} & \textbf{31.256 $\pm$ 0.241} & \textbf{31.322 $\pm$ 0.203}  & \multicolumn{1}{c|}{\textbf{31.321 $\pm$ 0.203}} & \textbf{16.021 $\pm$ 0.244} & \textbf{15.124 $\pm$ 0.244} & \textbf{15.642 $\pm$ 0.201} \\ 
\hline
\end{tabular}%
}
\end{minipage}

\begin{minipage}[ht]{0.95\linewidth}
\resizebox{\columnwidth}{!}{%
\begin{tabular}{ccccccc}
\hline
\multicolumn{1}{c|}{Dataset} &  \multicolumn{3}{c|}{Book-Crossing} & \multicolumn{3}{c}{GoodReads} \\ \hline
\multicolumn{1}{c|}{Measure (\%)} & \multicolumn{1}{c|}{Recall} & \multicolumn{1}{c|}{Precision} & \multicolumn{1}{c|}{nDCG} & \multicolumn{1}{c|}{Recall} & \multicolumn{1}{c|}{Precision} & nDCG \\ \hline
\multicolumn{1}{c|}{DDPG}&  7.442 $\pm$ 0.452 & 5.315 $\pm$ 0.232 & \multicolumn{1}{c|}{5.824 $\pm$ 0.281} & 11.652 $\pm$ 0.188 & 10.246 $\pm$ 0.142 & 9.941 $\pm$ 0.189\\ 
\multicolumn{1}{c|}{SAC} &  7.211 $\pm$ 0.293 &  5.114 $\pm$ 0.390 & \multicolumn{1}{c|}{5.724 $\pm$ 0.226} & 11.452 $\pm$ 0.242 & 10.027 $\pm$ 0.181 & 9.721 $\pm$ 0.220\\ 
\multicolumn{1}{c|}{TD3} & 7.025 $\pm$ 0.193 & 5.105 $\pm$ 0.179 & \multicolumn{1}{c|}{5.665 $\pm$ 0.332} & 11.028 $\pm$ 0.199 & 9.876 $\pm$ 0.167 & 9.521 $\pm$ 0.246\\  
\multicolumn{1}{c|}{DT} &  8.375 $\pm$ 0.322 & 6.892 $\pm$ 0.188 & \multicolumn{1}{c|}{7.587 $\pm$ 0.228}& 12.957 $\pm$ 0.177 & 10.998 $\pm$ 0.169 & 10.291 $\pm$ 0.206\\  
\multicolumn{1}{c|}{DT4Rec} & 8.667 $\pm$ 0.221 & 6.984 $\pm$ 0.149 & \multicolumn{1}{c|}{7.833 $\pm$ 0.189} & 12.972 $\pm$ 0.189 & 10.994 $\pm$ 0.191 & 10.425 $\pm$ 0.298\\  
\multicolumn{1}{c|}{CDT4Rec} & 9.234 $\pm$ 0.123 & 7.226 $\pm$ 0.289 & \multicolumn{1}{c|}{8.276 $\pm$ 0.279} & 13.274 $\pm$ 0.287 & 11.276 $\pm$ 0.175 & 10.768 $\pm$ 0.372  \\ 
\hline
\multicolumn{1}{c|}{EDT4Rec} & \textbf{10.345 $\pm$ 0.102} & \textbf{8.254 $\pm$ 0.176} & \multicolumn{1}{c|}{\textbf{9.141 $\pm$ 0.234}} & \textbf{14.144 $\pm$ 0.201} & \textbf{12.422 $\pm$ 0.145} & \textbf{11.641 $\pm$ 0.245}   \\ 
\hline
\end{tabular}%
}
\end{minipage}

\begin{minipage}[ht]{0.95\linewidth}
\resizebox{\columnwidth}{!}{%
\begin{tabular}{ccccccc}
\hline
\multicolumn{1}{c|}{Dataset} &  \multicolumn{3}{c|}{MovieLens-20M} & \multicolumn{3}{c}{Netflix} \\ \hline
\multicolumn{1}{c|}{Measure (\%)} & \multicolumn{1}{c|}{Recall} & \multicolumn{1}{c|}{Precision} & \multicolumn{1}{c|}{nDCG} & \multicolumn{1}{c|}{Recall} & \multicolumn{1}{c|}{Precision} & nDCG \\ \hline
\multicolumn{1}{c|}{DDPG} & 17.622 $\pm$ 0.251 & 15.889 $\pm$ 0.241 &  \multicolumn{1}{c|}{15.802 $\pm$ 0.201} & 13.305 $\pm$ 0.235 & 11.788 $\pm$ 0.213 & 10.834 $\pm$ 0.214\\ 
\multicolumn{1}{c|}{SAC}& 17.782 $\pm$ 0.204 & 15.989 $\pm$ 0.229 &  \multicolumn{1}{c|}{15.821 $\pm$ 0.212} & 13.324 $\pm$ 0.210 & 11.981 $\pm$ 0.246 & 10.970 $\pm$ 0.270\\ 
\multicolumn{1}{c|}{TD3}&  17.531 $\pm$ 0.291 & 15.866 $\pm$ 0.261 &  \multicolumn{1}{c|}{15.731 $\pm$ 0.302} & 13.231 $\pm$ 0.284 & 11.676 $\pm$ 0.204 & 10.789 $\pm$ 0.266\\  
\multicolumn{1}{c|}{DT}  & 18.889 $\pm$ 0.224 & 16.989 $\pm$ 0.242 &  \multicolumn{1}{c|}{16.912 $\pm$ 0.202} & 14.598 $\pm$ 0.202 & 12.251 $\pm$ 0.141 & 11.864 $\pm$ 0.242\\  
\multicolumn{1}{c|}{DT4Rec}  & 18.971 $\pm$ 0.254 & 17.013 $\pm$ 0.289 &  \multicolumn{1}{c|}{16.942 $\pm$ 0.248} & 14.879 $\pm$ 0.254 & 12.786 $\pm$ 0.198 & 11.961 $\pm$ 0.261\\  
\multicolumn{1}{c|}{CDT4Rec} & 19.273 $\pm$ 0.212 & 17.371 $\pm$ 0.276 &  \multicolumn{1}{c|}{17.311 $\pm$ 0.216} & 15.271 $\pm$ 0.127 & 13.274 $\pm$ 0.168 & 12.479 $\pm$ 0.198\\  
\hline
\multicolumn{1}{c|}{EDT4Rec} & \textbf{20.314 $\pm$ 0.201} & \textbf{18.341 $\pm$ 0.221} &  \multicolumn{1}{c|}{\textbf{18.234 $\pm$ 0.186}} & \textbf{16.541 $\pm$ 0.201} & \textbf{14.356 $\pm$ 0.181} & \textbf{13.661 $\pm$ 0.145}\\  
\hline
\end{tabular}%
}
\end{minipage}
\label{tab:result}
\end{table*}
In our study, we focus on evaluating EDT4Rec within the context of DT-based methods. Most existing works in RLRS have been assessed using customized settings, which makes fair comparison challenging~\cite{chen2023deep}. To address this, we have selected a range of baselines, encompassing both prominent DT-based methods and well-established RL algorithms, for a comprehensive evaluation:
\begin{itemize}
    \item \textbf{Deep Deterministic Policy Gradient (DDPG)}~\cite{lillicrap2015continuous}: An off-policy method designed for environments with continuous action spaces.
    
    \item \textbf{SAC}~\cite{haarnoja2018soft}: This approach is an off-policy, maximum entropy Deep RL method that focuses on optimizing a stochastic policy. While we are using deterministic policy here.

    \item \textbf{Twin Delayed DDPG (TD3)}~\cite{fujimoto2018addressing}: An enhancement of the baseline DDPG, TD3 improves performance by learning two Q-functions, updating the policy less frequently.

    \item \textbf{DT}~\cite{chen2021decision}: An offline RL algorithm that leverages the transformer architecture to infer actions.

    \item \textbf{DT4Rec}~\cite{zhao2023user}: Building on the standard DT framework, DT4Rec integrates a conservative learning method to better understand users' intentions in offline RLRS.

    \item \textbf{CDT4Rec}~\cite{Wang_2023}: This model introduces a causal layer to the DT framework, aiming to more effectively capture user intentions and preferences in offline RLRS.
\end{itemize}

\begin{figure*}[!ht]
     \centering
     \begin{subfigure}[b]{0.33\linewidth}
         \centering
           \includegraphics[width=\linewidth]{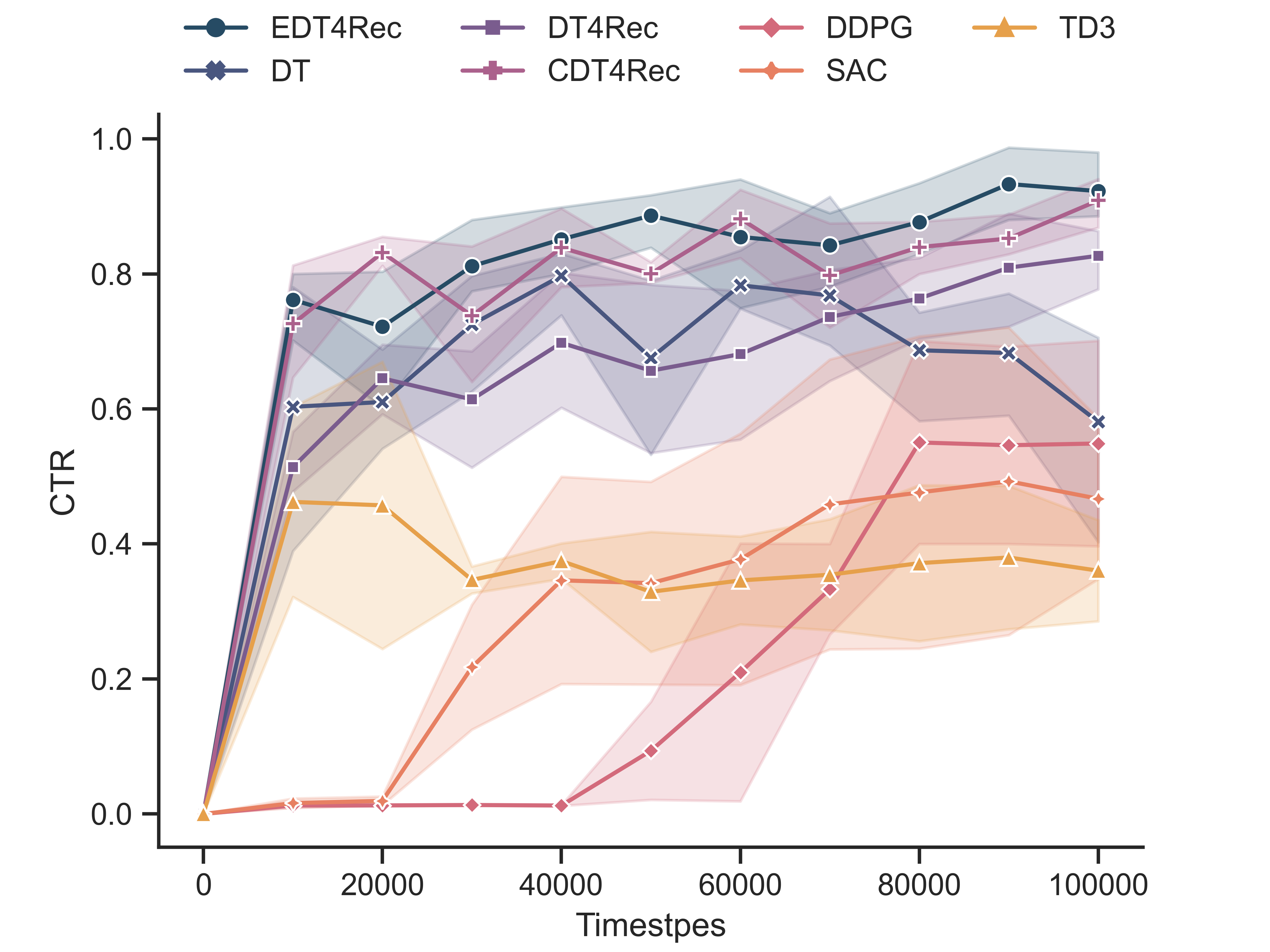}
           \caption{}
           \label{fig:over_comp}
     \end{subfigure}
     \begin{subfigure}[b]{0.33\linewidth}
         \centering
         \includegraphics[width=\linewidth]{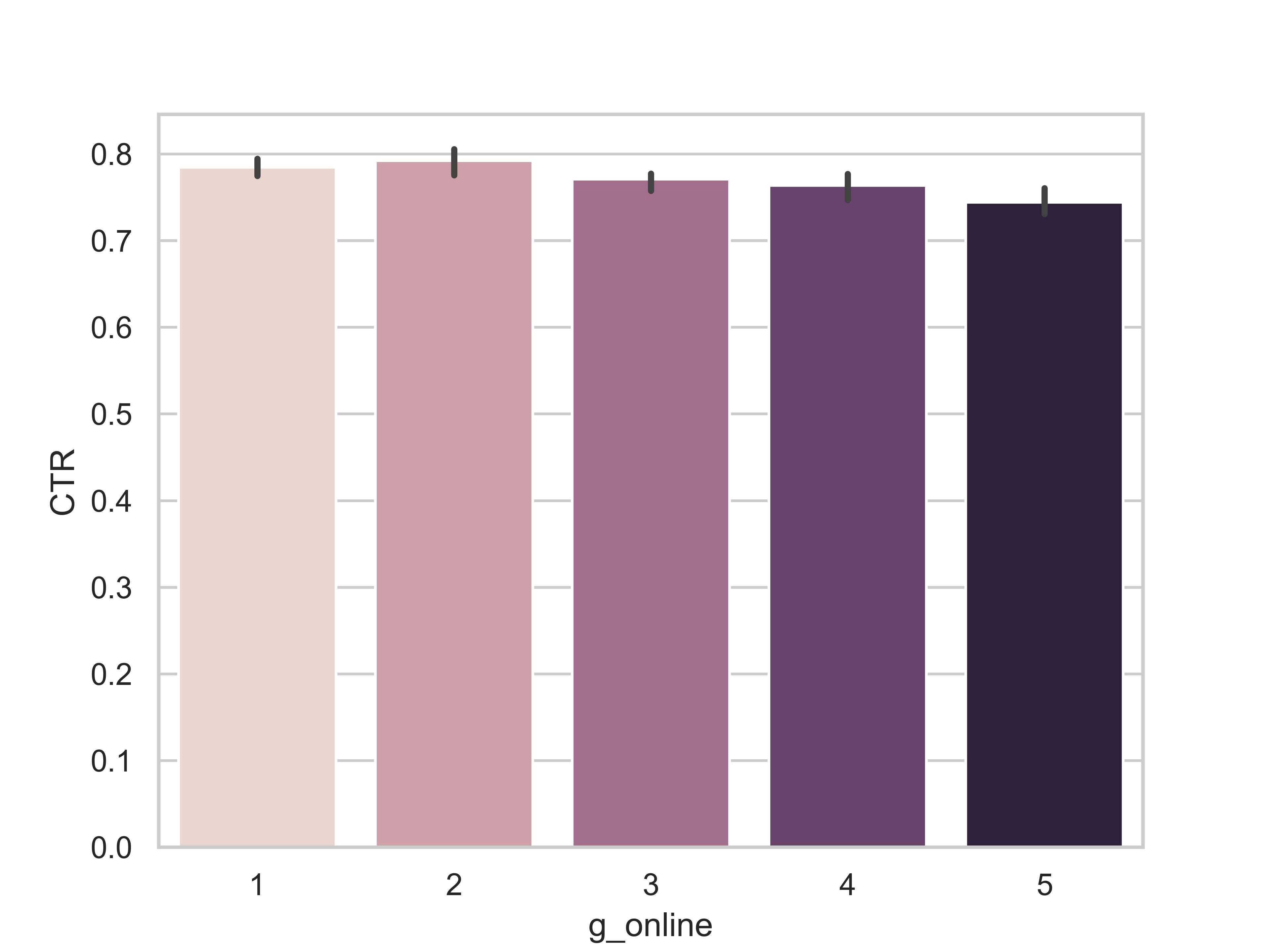}
         \caption{}
         \label{fig:g_online}
     \end{subfigure}
     \begin{subfigure}[b]{0.33\linewidth}
         \centering
         \includegraphics[width=\linewidth]{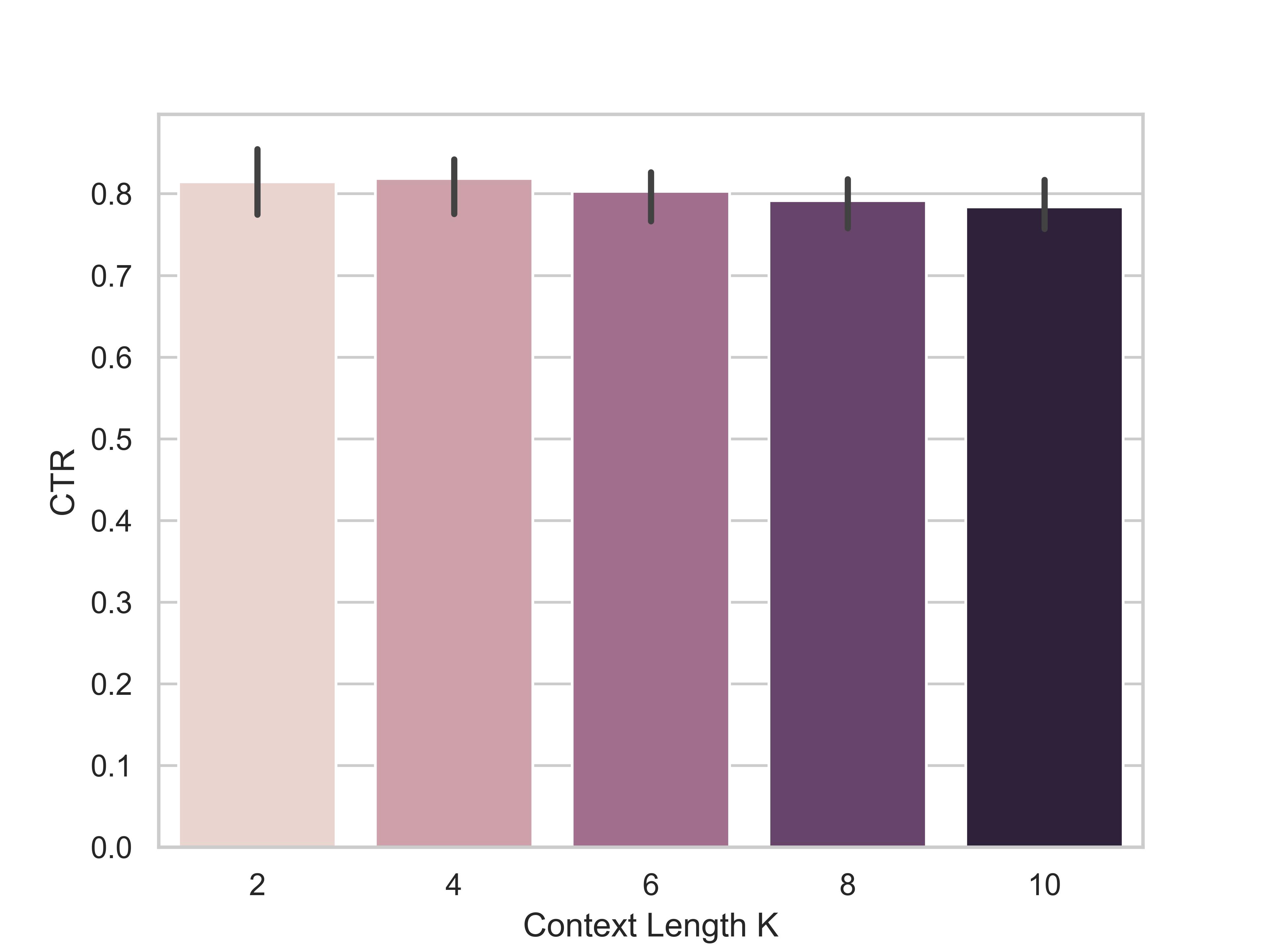}
         \caption{}
         \label{fig:k}
     \end{subfigure}
        \caption{(a). Overall comparison result with variance between the baselines and EDT4Rec in the VirtualTaobao simulation environment. (b).Hyperparameter $g_{online}$ Study, the value reported in the average CTR over $100,000$ timesteps. (c).Hyperparameter $K$ Study, the value reported in the average CTR over $100,000$ timesteps}
\end{figure*}

\subsection{Overall Comparison (RQ1)}
The summarized results of our experiments with the offline dataset are presented in~\Cref{tab:result}. Notably, EDT4Rec demonstrates superior performance over all baseline methods, including those DT based offline RLRS and traditional RL approaches. A key observation is the general superiority of DT-based methods over RL-based methods, likely attributable to the transformer framework's expressiveness.

In our online simulator experiments, EDT4Rec was compared against the baselines mentioned earlier. The comparative results, as detailed in~\Cref{fig:over_comp}, reveal a significant improvement by EDT4Rec, particularly in the later timesteps. Moreover, we observe that DT-based methods outperform traditional RL algorithms. The reason would be those methods accessing more recorded offline trajectories and benefiting from the transformer's high expressiveness.

However, it's observed that DT, DT4Rec, and CDT4Rec perform worse than EDT4Rec, given that the pre-collected data are sub-optimal. This observation reinforces our assertion that DT lacks exploration capability and that its pre-training stage closely resembles behavior cloning.

\subsection{Hyperparameter Study (RQ2)}
In this section, we delve into the hyperparameter analysis for our proposed EDT4Rec model, particularly examining the impacts of $\vecg_{online}$ and $K$. The hyperparameter $\vecg_{online}$, set at 2 in our experiments, plays a crucial role in regulating exploration. Considering that the maximum reward per interaction is capped (e.g., at 10 in the VirtualTB environment), a smaller $\vecg_{online}$ value is preferable to encourage exploration over exploitation during the fine-tuning phase. The results, illustrating the performance of EDT4Rec with varying $\vecg_{online}$, are presented in~\Cref{fig:g_online}. We observe that the optimal performance is achieved when $\vecg_{online}$ is set to 2.
Turning our attention to the context length $K$, this parameter determines the number of previous actions accessible to the RTG. Following insights from previous work~\cite{Wang_2023} that suggest a smaller context length is beneficial, we explore a range of $K$ values: $\{2,4,6,8,10\}$, while keeping $\vecg_{online}$ fixed at 2. Similar to $\vecg_{online}$, the average CTR over 100,000 timesteps for different $K$ values is reported in~\Cref{fig:k}. It becomes evident that the proposed EDT4Rec model reaches its peak performance when both $K$ and $\vecg_{online}$ are set to 2.
\subsection{Ablation Study (RQ3)}
\begin{figure}[h]
    \centering
    \includegraphics[width=0.8\linewidth]{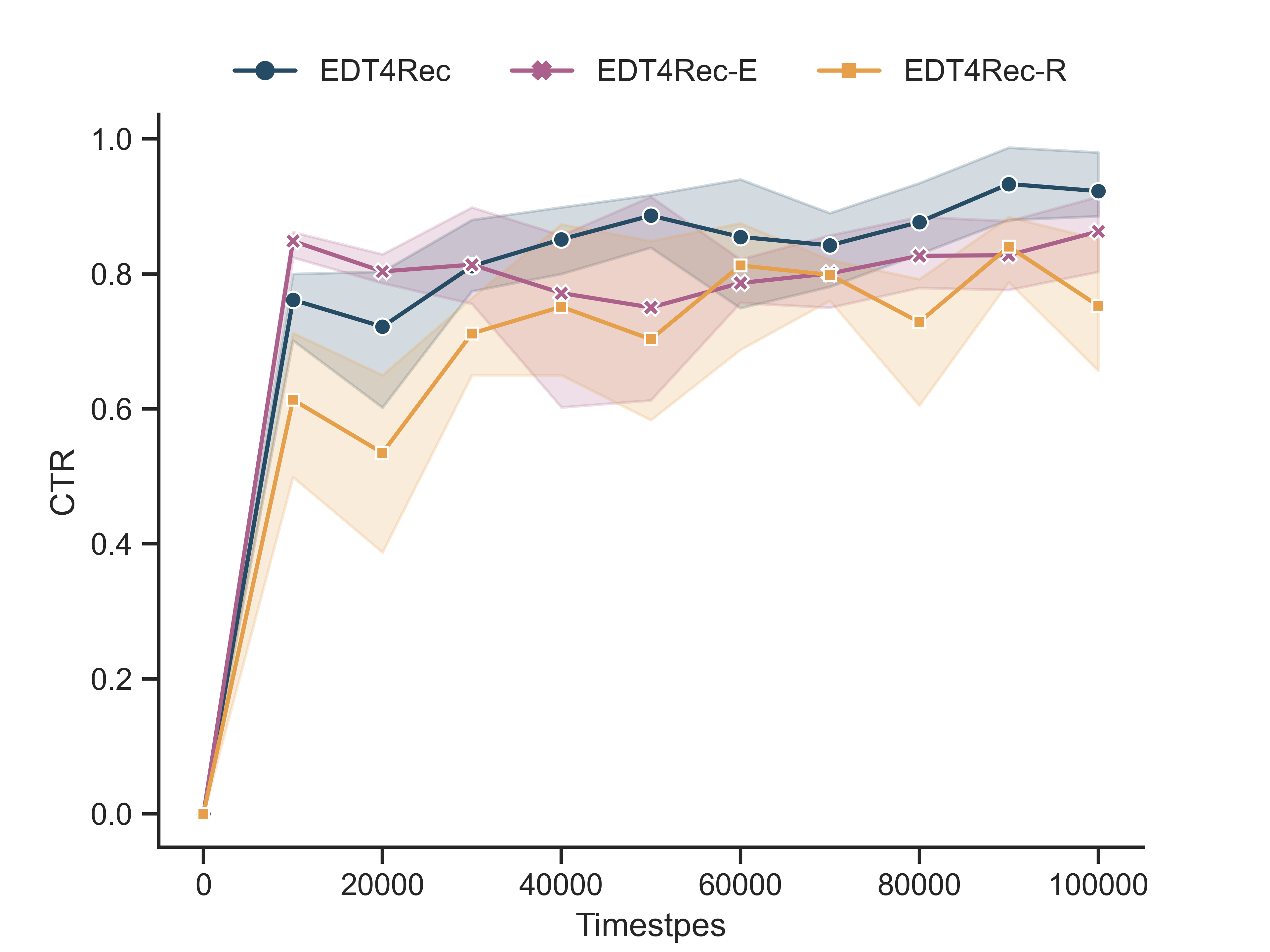}
    \caption{Ablation Study}
    \label{fig:ablation-label}
\end{figure}

In order to comprehensively understand the impact of exploration and reward relabeling on the proposed EDT4Rec model, we conducted an ablation study. This study isolates the effects of these components by evaluating two variations of EDT4Rec: EDT4Rec-E, which excludes the online exploration goal, and EDT4Rec-R, which omits the reward relabeling feature.
The results of this study, shows in~\Cref{fig:ablation-label}, provides some insightful trends. Notably, in the absence of the exploration component (EDT4Rec-E), the model initially outperforms the complete EDT4Rec. This could be attributed to EDT4Rec-E solely leveraging exploitation based on previously learned policies without engaging in exploration. However, as the agent encounters unfamiliar states over time, the lack of exploration leads to a decline in performance due to the inadequacy of the existing policy in making accurate recommendations in these new scenarios.
In addition, the removal of the reward relabeling component in EDT4Rec-R results in a significant drop in performance. This outcome underscores the critical role of online exploration in the effectiveness of EDT4Rec. The inability to adaptively relabel rewards based on online experiences evidently hampers the model's ability to effectively navigate and learn from the environment, thereby highlighting the essential contribution of each component to the overall success of the EDT4Rec model. It also support out claim about that DT lacks the capability of stitching.

\section{Conclusion}
In this study, we introduced Max-Entropy enhanced Decision Transformer with Reward Relabeling for Offline RLRS (EDT4Rec), a novel model crafted to overcome two significant challenges in Decision Transformer (DT)-based systems: i) the lack of stitching capability, and ii) insufficient online exploration, which can lead to sub-optimal performance in scenarios with sub-optimal datasets. Our comprehensive experiments demonstrate that EDT4Rec surpasses existing DT-based methods in performance.
Looking ahead, our future work will delve deeper into the stitching problem. Although the current reward relabeling strategy in EDT4Rec relies on learned value functions from optimal trajectories, such trajectories may not always be available. Therefore, we aim to develop a more straightforward method of reward relabeling, one that does not necessitate reliance on optimal trajectories. This advancement will further enhance the applicability and robustness of EDT4Rec in diverse real-world scenarios.
\bibliographystyle{ACM-Reference-Format}
\balance
\bibliography{sample-base}

\end{document}